\documentclass[%
reprint,
amsmath,amssymb,
prab,
]{revtex4-2}
\usepackage{tikz}
\usetikzlibrary{shapes,arrows.meta,positioning,calc}
\usepackage{graphicx}
\usepackage{dcolumn}
\usepackage{bm}
\usepackage[colorlinks=true,linkcolor=blue,urlcolor=blue,citecolor=blue]{hyperref}
\usepackage{siunitx}
\usepackage{booktabs}
\usepackage{array}
\usepackage{multirow}

\begin{document}
	
	\title{Reactive feedback for suppression of the periodic transient beam loading effect in storage rings: theory and simulations}
	
	\author{PeiZhi Fang}
	\affiliation{Department of Engineering Physics, Tsinghua University, Beijing 100084, China}
	\author{Murilo B. Alves}
	\affiliation{MAX IV Laboratory, Lund University, Lund, Sweden}
	\author{Hongjuan Yao}
	\affiliation{Department of Engineering Physics, Tsinghua University, Beijing 100084, China}
	\author{Shuxin Zheng}
	\email{zhengsx@tsinghua.edu.cn}
	\affiliation{Department of Engineering Physics, Tsinghua University, Beijing 100084, China}
	\author{Chuanxiang Tang}
	\affiliation{Department of Engineering Physics, Tsinghua University, Beijing 100084, China}
	\date{\today}
	
	\begin{abstract}
		\noindent In fourth-generation synchrotron light sources, harmonic cavities are often employed for bunch lengthening. The reduced longitudinal focusing, combined with high total $R/Q$ of these cavities, can make the beam susceptible to the periodic transient beam loading (PTBL) effect, a multibunch instability characterized by slow oscillations in bunch centroids and lengths and by its resistance to conventional feedback damping. In this work, building on a recent proposal of applying reactive feedback to control the PTBL effect, we developed a frequency-domain theoretical model, including the feedback voltage in the linearized Vlasov equation to derive a Lebedev equation modified by the feedback transfer function. The reactive feedback was also implemented in the time domain in the macroparticle tracking code \texttt{STABLE}. The methods were applied to the parameters of the double-rf system of the MAX IV storage ring. Both approaches show that, above a threshold gain, an idealized pure reactive feedback is able to suppress the PTBL effect, with good agreement between theory and simulations across beam currents and $R/Q$ values. Realistic hardware limitations were included in the analysis, determining the limits for feedback gain, latency, and bandwidth, which are compatible with existing bunch-by-bunch feedback systems. Using the HALF storage ring parameters, the semi-analytical results indicate that the reactive feedback can be extended to suppress the PTBL effect in triple-rf systems, albeit at higher threshold gains than for double-rf systems.
	\end{abstract}
	
	\maketitle
	
	\section{Introduction}
	Fourth-generation synchrotron light sources utilize ultra-low transverse emittances to deliver high-brightness radiation to the beamlines. The high electron density intensifies intrabeam scattering (IBS), which blows up the beam emittance and energy spread~\cite{KuboMtingwaWolski-2005}, and shortens the Touschek lifetime~\cite{AlvesAnderssonCullinan-2024}. To maintain adequate lifetime and alleviate collective instabilities, harmonic cavities (HCs) are typically installed in storage rings to lengthen electron bunches \cite{ByrdGeorgsson-6278,HeLi-6377,Cullinan-6394,YamamotoSakanaka-6383}.
	
	Nevertheless, lengthened bunches become highly susceptible to voltage fluctuations in the HCs. As the beam passes through the cavities, the beam-induced voltage distorts the longitudinal bunch profile; in turn, this deformation of the bunch profile modulates the amplitude and phase of the induced voltage \cite{He-6289}. Once this mutual coupling forms a positive feedback loop, it excites the Periodic Transient Beam Loading (PTBL) effect \cite{HeLi-6145,He-6289}.
	This effect drives macroscopic oscillations in both the bunch centroid and bunch length, affecting the stable operation of the storage ring. Recently, the PTBL effect has drawn extensive attention in simulation studies at many light sources, such as Diamond-II~\cite{Olsson-6385}, SOLEIL-II~\cite{Gamelin-6293}, SIRIUS~\cite{Alves-6277}, and HALF~\cite{HeLi-6145,He-6289}, and has been observed in experiments at MAX IV \cite{CullinanAndersson-6290}.
	
	To mitigate the PTBL effect, recent efforts have primarily focused on lowering the $R/Q$ of harmonic cavities, prompting facilities such as SOLEIL-II~\cite{Gamelin-6293}, KEK-LS~\cite{Yamamoto-6380}, and HALF~\cite{He-6289} to adopt $\mathrm{TM020}$ cavities or superconducting technology. Alternatively, direct rf feedback has been investigated to reduce the effective shunt impedance seen by the beam, indirectly lowering the apparent $R/Q$ of the cavity, but simulation studies show that it extends stable operation only slightly beyond the PTBL threshold \cite{YamamotoYamaguchi-6382}.
	
	Bunch-by-bunch (BbB) feedback systems are standard tools for suppressing longitudinal coupled-bunch instabilities, conventionally operating in resistive mode to provide damping to the beam. However, such resistive feedback provides little damping against the PTBL effect \cite{Lindberg-6386}. Recently, Ref.~\cite{Alves-6384} described the mechanism of PTBL as loss of coherent focusing for coupled-bunch modes and proposed reactive feedback as a potential mitigation scheme. Nevertheless, a theoretical formulation and a quantitative assessment of the efficacy of this approach have yet to be established.
	
	
	The paper is organized as follows: In Sec. \ref{sec:principles}, we introduce the concept of reactive feedback and qualitatively elucidate its stabilization mechanism against the PTBL effect from time-domain and frequency-domain perspectives. In Sec. \ref{sec:theory}, we establish the theoretical framework by incorporating the feedback term into the longitudinal Vlasov equation, deriving the Lebedev equation and formulating the effective synchrotron frequency method. In Sec. \ref{sec:tracking}, macroparticle tracking simulations based on the \texttt{STABLE} code are conducted to verify the theoretical model. We present time-domain damping dynamics, threshold gain benchmarks, and the influence of the main cavity voltage, evaluate critical engineering challenges including loop latency and cavity bandwidth, and extend the analysis to the triple-rf system of the HALF storage ring. Finally, conclusions and future outlook are summarized in Sec. \ref{sec:summary}.

	\section{Reactive Feedback: Concept and Stabilization Mechanism}
	\label{sec:principles}
	
	Before developing the analytical model, this section introduces the reactive feedback concept and qualitatively describes how it suppresses the PTBL effect.
	
	We consider the feedback as an ideal bunch-by-bunch controller acting on the beam centroid: on each turn, the centroid offset of each bunch is measured, and a correction voltage proportional to this offset is applied to that bunch.
	
	Unlike resistive feedback, which provides a quadrature kick to damp instabilities, reactive feedback adjusts the loop phase so that the corrective voltage is strictly in phase with the centroid displacement:
	\begin{equation}
		V_{\text{FB}}(t) = G \langle z \rangle(t), \label{eq:v_fb_intro}
	\end{equation}
	where $G$ denotes the reactive feedback gain.
	
	The stabilization mechanism of reactive feedback can be interpreted from both time-domain and frequency-domain perspectives.
	
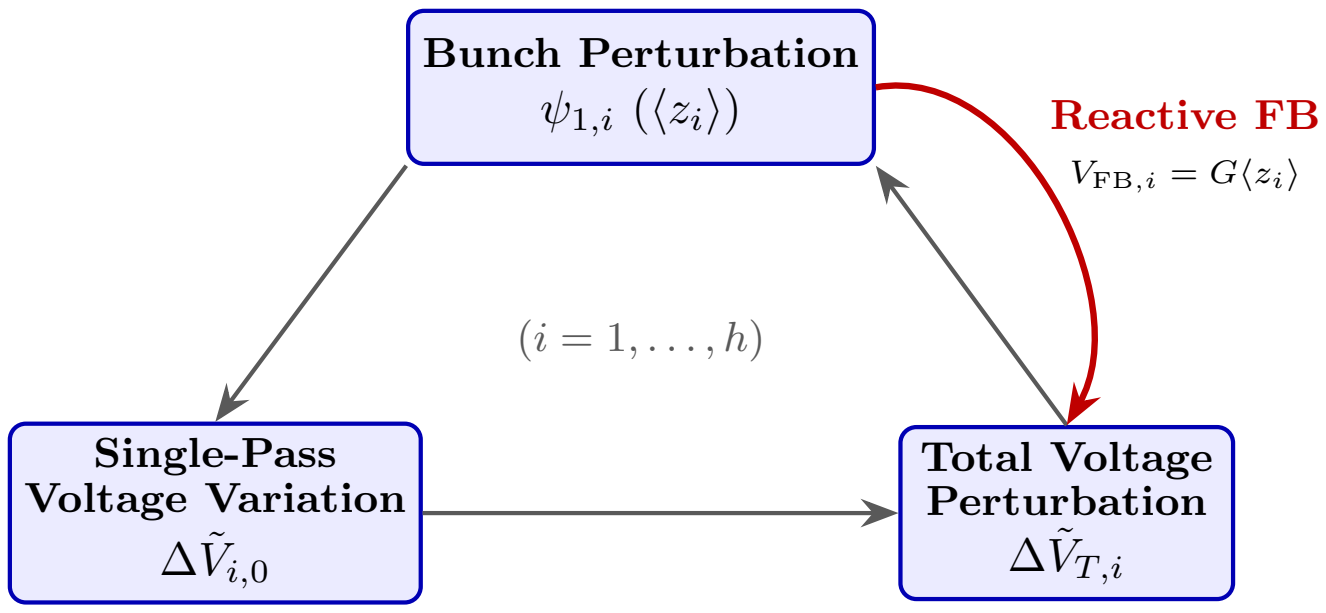
\begin{figure}[htbp]
		\centering
		\begin{tikzpicture}[
			scale=0.88, every node/.style={transform shape},
			box/.style={draw=blue!70!black, fill=blue!8, thick, rounded corners=3pt, minimum width=2.5cm, minimum height=1.15cm, align=center},
			arrow/.style={-{Stealth[length=2.5mm]}, thick, draw=gray!70!black},
			fbarrow/.style={-{Stealth[length=2.7mm]}, very thick, draw=red!75!black}
			]
			
			\node[box] (rho) at (0, 3.2) {\textbf{Bunch Perturbation}\\[1.5pt]\normalsize $\psi_{1,i} \ (\langle z_i \rangle)$};
			\node[box] (v0) at (-3.2, 0) {\textbf{Single-Pass}\\[-1pt]\textbf{Voltage Variation}\\[1.5pt]\normalsize $\Delta\tilde{V}_{i,0}$};
			\node[box] (vt) at (3.2, 0) {\textbf{Total Voltage}\\[-1pt]\textbf{Perturbation}\\[1.5pt]\normalsize $\Delta\tilde{V}_{T,i}$};
			
			\draw[arrow] (rho.south west) -- (v0.north);
			\draw[arrow] (v0.east) -- (vt.west);
			\draw[arrow] (vt.north) -- (rho.south east);
			
			\node[font=\small, text=gray!70!black] at (0, 1.3) {$(i = 1, \dots, h)$};
			
			\draw[fbarrow] (rho.east) to[out=10, in=60, looseness=0.95] 
			node[above right=-2pt and -4pt, align=center] {
				\small \textbf{\textcolor{red!75!black}{Reactive FB}}\\[1pt]
				\scriptsize $V_{\text{FB},i} = G\langle z_i \rangle$
			} (vt.north);
			
		\end{tikzpicture}
		\caption{Dynamical perturbation loop around the stationary equilibrium and the suppression mechanism introduced by the reactive feedback system.}
		\label{fig:self_consistent_loop}
	\end{figure}

	From a time-domain viewpoint, the PTBL effect originates from a positive feedback mechanism around the stationary equilibrium $\Psi_0$. A longitudinal perturbation $\Psi_1$ introduces bunch centroid displacements $\langle z_i \rangle$ relative to the equilibrium positions, which modulate the bunch form factors and the beam-induced voltages. The accumulated voltage perturbation $\Delta\tilde{V}_{T,i}$ across all bunches acts back on the beam to further distort the bunch distribution, driving a secondary position displacement $\langle z_i' \rangle$. Under heavy beam loading with high $R/Q$, when this secondary displacement exceeds the initial one ($\langle z_i' \rangle > \langle z_i \rangle$), the perturbation grows exponentially, manifesting as a growing PTBL effect \cite{He-6289}. As illustrated by the outer loop in Fig.~\ref{fig:self_consistent_loop}, the reactive feedback introduces an external corrective path that couples the bunch centroid displacement $\langle z_i \rangle$ directly to the total voltage. This corrective kick directly counteracts the induced voltage perturbation $\Delta\tilde{V}_{T,i}$, suppressing the positive feedback loop and thereby stabilizing the beam.
	
	From a frequency-domain perspective, the PTBL effect is driven predominantly by the reactive impedance of the harmonic cavity, whose positive detuning ($\omega_r > n_{\text{hc}}\omega_{\text{rf}}$, where $n_{\text{hc}}$ is the harmonic number of the HC) induces a strong negative coherent frequency shift in the coupled-bunch dipole mode ($\ell=1$) \cite{Venturini-6282, Alves-6384}.
	In the bunch-lengthening regime, this shift can overcome the weakened incoherent synchrotron frequency and drive the coherent mode frequency toward zero, such that the net restoring force on the coherent motion vanishes and the perturbation diverges. Because resistive feedback provides only damping, it cannot prevent this frequency collapse, and its damping capacity is easily exceeded by the resulting high growth rate.
	In contrast, reactive feedback applies an in-phase kick $V_{\text{FB}} = G \langle z \rangle$, which acts as an external restoring force that directly offsets the negative coherent frequency shift, keeping the coherent dipole frequency away from zero and thereby avoiding the growth of PTBL oscillations \cite{Alves-6384}.
	
	\section{Linearized Vlasov equation with feedback}\label{sec:theory}
	In this section, we will develop the theoretical basis for the semi-analytical methods applied to evaluate the impact of active feedback systems on coupled-bunch instabilities in general. The particular case of interest will be the PTBL effect driven by harmonic cavities in bunch-lengthening mode. 
	
	The two-dimensional phase-space distribution $\Psi$ in longitudinal action-angle variables $(J,\varphi)$ is governed by the Vlasov equation:
	\begin{equation}
		\frac{d\Psi}{ds} = \frac{\partial\Psi}{\partial s} + \{\Psi,\mathcal{H}\} = 0,
		\label{eq:vlasov}
	\end{equation}
	where the independent variable $s$ is the longitudinal position of the synchronous particle along the ring, $\{\Psi,\mathcal{H}\}$ denotes the Poisson bracket and $\mathcal{H}$ is the longitudinal Hamiltonian. We consider a small coherent perturbation oscillating at a complex frequency $\Omega$ around the equilibrium set by $\mathcal{H}_{0}(J)$ and $\Psi_{0}(J)$, which are obtained as solutions to the Haissinki equation and can include potential-well distortion effects due to wakefields:
	\begin{align}
		\mathcal{H}(J,\varphi,s) &= \mathcal{H}_{0}(J) + \Phi_{1}(J,\varphi) e^{-i\Omega s/c}, \label{eq:perturbed_hamiltonian}\\
		\Psi(J,\varphi,s) &= \Psi_{0}(J) + \Psi_{1}(J,\varphi) e^{-i\Omega s/c}, \label{eq:perturbed_dist}
	\end{align}
	where $\Psi_{0}(J) = (2\pi\mathcal{N}_{J})^{-1}\exp[-\mathcal{H}_{0}(J)/(\alpha_c\sigma_{\delta}^{2})]$ is the equilibrium distribution. The normalization factor $\mathcal{N}_{J}$ is such that $\int_{0}^{\infty}dJ \int_{0}^{2\pi} d\varphi\Psi_0 = 1$; $\alpha_c$ is the momentum compaction factor and $\sigma_{\delta}$ the relative energy spread.
	
	Linearizing Eq.~\eqref{eq:vlasov} to first order in the perturbations yields:
	\begin{equation}
		-i\Omega\Psi_{1} + \omega_{s}(J)\frac{\partial\Psi_{1}}{\partial\varphi} - c\frac{\partial\Psi_{0}}{\partial J}\frac{\partial\Phi_{1}}{\partial\varphi} = 0,
		\label{eq:linearized_vlasov}
	\end{equation}
	where $\omega_{s}(J) = c\frac{\partial\mathcal{H}_{0}}{\partial J}$ is the amplitude-dependent synchrotron frequency.
	
	In the presence of a feedback system, the perturbed potential $\Phi_{1}$ has two components: the beam-induced voltage due to wakefields and the active feedback voltage:
	\begin{align}
		\Phi_{1}(z) &= \Phi_{1}^{\text{wake}}(z) + \Phi_{1}^{\text{FB}}(z) \\ 
		&= -\frac{e}{E_0C_0}\int_{0}^{z} dz' {\left(\hat{V}_{\text{wake}}(z') + \hat{V}_{\text{FB}}\right)},
		\label{eq:phi1_total}
	\end{align}
	where $E_{0}$ is the nominal beam energy, and $C_{0}$ is the ring circumference. 
	
	We will focus on even filling cases with $M$ equally spaced bunches, and consider perturbations for a coupled-bunch mode number $\ell$. The wake voltage driven by the perturbed bunch spectrum $\tilde{\lambda}_{1}(\omega)$ is
	\begin{equation}
		\hat{V}_{\text{wake}}(z) = -I_{0}\sum_{p=-\infty}^{\infty}\tilde{\lambda}_{1}(\omega_{p,\ell})Z_{p,\ell}(\Omega)e^{-i\omega_{p,\ell}z/c},
		\label{eq:v_wake}
	\end{equation}
	where $I_{0}$ is the average beam current, $\omega_{0}$ is the revolution angular frequency, $\omega_{p,\ell} = (pM + \ell)\omega_{0}$ are the coupled-bunch sideband frequencies, and $Z_{p,\ell}(\Omega) = Z_{\parallel}(\omega_{p,\ell} + \Omega)$ denotes the longitudinal impedance.
	
	Assuming that the feedback system is linear, time-invariant and acts on the beam centroid, the generated feedback voltage can be expressed in the frequency domain as
	\begin{equation}
		\hat{V}_{\text{FB}}(\omega) = T_{\text{FB}}(\omega)\tilde{\langle{z}\rangle}({\omega}),
		\label{eq:v_fb}
	\end{equation}
	where $T_{\text{FB}}(\omega)$ is the feedback transfer function. From the assumption of a coherent perturbation oscillating at frequency $\Omega$, the only Fourier component of the dynamical centroid displacement is at frequency $\Omega$, given by
	\begin{equation}
		\tilde{\langle{z}\rangle}_\Omega = \int_{0}^{\infty}dJ \int_{0}^{2\pi}d\varphi\,\zeta(J,\varphi)\Psi_{1}(J,\varphi),
		\label{eq:centroid_def}
	\end{equation}
	where the canonical transformation $z = \zeta(J,\varphi)$ was considered. With that, the feedback voltage is $\hat{V}_{\text{FB}}(\Omega) = T_{\text{FB}}(\Omega)\tilde{\langle{z}\rangle}_\Omega$ and its longitudinal potential is simply $\Phi_{1}^{\text{FB}}(z) = -e\hat{V}_{\text{FB}}z/E_{0}C_{0}$.
	
	Taking the derivative of the potentials with respect to $\varphi$, the wakefield and feedback contributions are
	\begin{align}
		\frac{\partial\Phi_{1}^{\text{wake}}}{\partial\varphi} &= \frac{eI_{0}}{E_{0}C_{0}}\sum_{p=-\infty}^{\infty}\frac{\tilde{\lambda}_{1}(\omega_{p,\ell})Z_{p,\ell}(\Omega)}{-i\omega_{p,\ell}/c}\frac{\partial}{\partial\varphi}e^{-i\omega_{p,\ell}\zeta(J,\varphi)/c},
		\label{eq:phi1_wake_deriv} \\
		\frac{\partial\Phi_{1}^{\text{FB}}}{\partial\varphi} &= -\frac{eT_{\text{FB}}(\Omega)\tilde{\langle z\rangle}_{\Omega}}{E_{0}C_{0}}\frac{\partial\zeta(J,\varphi)}{\partial\varphi}.
		\label{eq:phi1_fb_deriv}
	\end{align}
	
	To solve the linearized Vlasov equation~\eqref{eq:linearized_vlasov}, we expand the perturbation $\Psi_{1}(J,\varphi)$ in azimuthal modes $m$:
	\begin{equation}
		\Psi_{1}(J,\varphi) = \sum_{m\neq0} R_{m}(J) e^{im\varphi},
		\label{eq:psi1_fourier}
	\end{equation}
	and the canonical transformation $\zeta(J,\varphi)$ in Fourier series:
	\begin{equation}
		\zeta(J,\varphi) = \sum_{\nu=-\infty}^{\infty} \hat{f}_{\nu}(J) e^{i\nu\varphi}.
		\label{eq:z_fourier}
	\end{equation}
	Under these expansions, the perturbation line density spectrum $\tilde{\lambda}_{1}(\omega)$ and the dynamical centroid component can be written as:
	\begin{align}
		Y_p := \frac{\tilde{\lambda}_{1}(\omega_{p,\ell})}{2\pi} &= \sum_{m\neq0}\int_{0}^{\infty}dJ\, R_{m}(J) H_{m,p}(J),
		\label{eq:lambda_expansion}\\
		U := \frac{\tilde{\langle{z}\rangle}_\Omega}{2\pi} &= \sum_{m\neq0}\int_{0}^{\infty}dJ\, R_{m}(J)\hat{f}_{m}^{*}(J) \label{eq:centroid_expansion}
	\end{align}
	where $H_{m,p}(J)$ is a function defined as:
	\begin{equation}
		H_{m,p}(J) := \frac{1}{2\pi}\int_{0}^{2\pi}d\varphi\, e^{im\varphi + i\omega_{p,\ell}\zeta(J,\varphi)/c},
		\label{eq:lebedev_func}
	\end{equation}
	and we have also introduced the auxiliary quantities $Y_p$ and $U$.
	
	Hence, Eq.~\eqref{eq:linearized_vlasov} for each azimuthal mode $m$ can be written as:
	\begin{equation}
		\begin{aligned}
			\relax&0=[\Omega - m\omega_{s}(J)] R_{m}(J) + m\frac{\partial\Psi_{0}}{\partial J} \\ &\times \Bigg[ i\kappa\sum_{p=-\infty}^{\infty}\frac{Z_{p,\ell}(\Omega)}{\omega_{p,\ell}}H_{m,p}^{*}(J)Y_{p} 
			+ \kappa_{\text{FB}}(\Omega)\hat{f}_{m}(J)U \Bigg] ,
		\end{aligned}
		\label{eq:Vlasov_Y_U}
	\end{equation}
	where we have introduced the intensity constant $\kappa$ and the feedback coupling factor defined by
	\begin{align}
		\kappa &= \frac{2\pi e I_{0} c}{E_{0} C_{0}}, \\
		\kappa_{\text{FB}}(\Omega) &= \frac{\kappa}{I_{0}}T_{\text{FB}}(\Omega).
		\label{eq:kappa_fb}
	\end{align}
	
	\subsection{Lebedev equation with feedback}
	For convenience, we introduce a bilinear symmetric product $\mathcal{P}[A, B](\Omega)$ applied to sets of functions $A = \left\{A_m\right\}_{m\neq0}$ and $B = \left\{B_m\right\}_{m\neq0}$:
	\begin{align}
		\mathcal{P}[A,B](\Omega) &:= \sum_{m\neq0}\int_{0}^{\infty}dJ\,\mathcal{W}_{m}(J,\Omega)A_{m}(J)B_{m}(J),
		\label{eq:bilinear_form}
	\end{align}
	with the weighting function defined as
	\begin{equation}
		\mathcal{W}_{m}(J,\Omega)=\frac{m}{\Omega-m\omega_{s}(J)}\frac{\partial\Psi_{0}}{\partial J}. \label{eq:kernel_def}
	\end{equation}
	
	Using these definitions, we isolate $R_m(J)$ from Eq.~\eqref{eq:Vlasov_Y_U} and insert it into the definitions of $Y_p$, Eq.~\eqref{eq:lambda_expansion}, and $U$, Eq.~\eqref{eq:centroid_expansion}. This results in two coupled equations
	\begin{align}
		Y_{p'} + i\kappa\sum_{p=-\infty}^{\infty}\frac{Z_{p,\ell}(\Omega)}{\omega_{p,\ell}}G_{pp'}(\Omega)Y_{p} + \kappa_{\text{FB}}(\Omega)\xi_{p'}(\Omega)U &= 0, 		\label{eq:coupled_Y} \\
		\left[1 + \kappa_{\text{FB}}(\Omega)S_{\text{FB}}(\Omega)\right]U + i\kappa\sum_{p=-\infty}^{\infty}\frac{Z_{p,\ell}(\Omega)}{\omega_{p,\ell}}\eta_{p}(\Omega)Y_{p} &= 0,
		\label{eq:coupled_U}
	\end{align}
	where the elements introduced above can be calculated via the product $\mathcal{P}$:
	\begin{subequations}
		\begin{align}
			G_{pp'}(\Omega) &= \mathcal{P}[H_{m,p'}, H_{m,p}^{*}](\Omega), \label{eq:G_mat} \\
			\xi_{p}(\Omega) &= \mathcal{P}[\hat{f}_{m}, H_{m,p}](\Omega), \label{eq:xi_vec}  \\
			\eta_{p}(\Omega) &= \mathcal{P}[\hat{f}_{m}^{*}, H_{m,p}^{*}](\Omega), \label{eq:eta_vec} \\
			S_{\text{FB}}(\Omega) &= \mathcal{P}[\hat{f}_{m}, \hat{f}_{m}^{*}](\Omega). \label{eq:S_fb}
		\end{align}
	\end{subequations}
	
	Solving for the scalar amplitude $U$ from Eq.~\eqref{eq:coupled_U} and substituting it into Eq.~\eqref{eq:coupled_Y} yields the Lebedev equation with feedback:
	\begin{align}
		\sum_{p=-\infty}^{\infty}\left\{\delta_{pp'} + i\kappa\frac{Z_{p,\ell}(\Omega)}{\omega_{p,\ell}}\left[G_{pp'}(\Omega) + \Delta G_{pp'}^{\text{FB}}(\Omega)\right]\right\}Y_{p} = 0.
		\label{eq:lebedev_final}
	\end{align}
	The active feedback system modifies the passive Lebedev matrix through the rank-one modification:
	\begin{equation}
		\Delta G_{pp'}^{\text{FB}}(\Omega) = -\frac{\kappa_{\text{FB}}(\Omega)}{1 + \kappa_{\text{FB}}(\Omega)S_{\text{FB}}(\Omega)}\,\xi_{p'}(\Omega)\eta_{p}(\Omega).
		\label{eq:delta_G_fb}
	\end{equation}
	
	After truncating the azimuthal modes to $|m| \leq m_{\mathrm{max}}$ and selecting the harmonics $p$, Eq.~\eqref{eq:lebedev_final} becomes a finite matrix equation, with non-trivial solutions for the coherent mode frequencies $\Omega$, found as roots of the matrix determinant. The coherent oscillation frequency is $\Re(\Omega)$ and an instability is predicted if the growth rate $\Im(\Omega)$ exceeds $\tau_{\mathrm{rad}}^{-1}$, where $\tau_{\mathrm{rad}}$ is the radiation damping time of the longitudinal motion.
	
	\subsection{Effective constant synchrotron frequency}
	A simplified model can be established by replacing the amplitude-dependent synchrotron frequency, $\omega_s(J)$, by an effective constant value, $\overline{\omega}_s$. Note that this process removes the intrabunch frequency spread; therefore, Landau damping effects are absent in the simplified model. The relation ${\omega}_s = \alpha_c c\sigma_\delta/\sigma_z$ has proven useful for the choice of $\overline{\omega}_s$, even for cases where bunch lengthening is provided by harmonic cavities~\cite{Alves-6384, GamelinGubaidulin-6392}. Here, $\sigma_z$ always refers to the second central moment of the equilibrium longitudinal distribution $\lambda_{0}(z)$, obtained as the solution of the Haissinski equation.
	
	Furthermore, we make the approximation $\Omega \approx m\overline{\omega}_s$ for the arguments of both the impedance $Z_{p,\ell}$ and the feedback coupling factor $\kappa_{\mathrm{FB}}$. 
	
	Applying these modifications to Eq.~\eqref{eq:Vlasov_Y_U}, we can isolate the radial function $R_m(J)$. We then define $X_{m,p}$ and $V_m$ as the constituent elements of $Y_p = \sum_{m} X_{m,p}$ and $U = \sum_{m} V_m$. Applying the $R_m(J)$ expression to the definitions of $X_{m,p}$ and $V_m$ yields the following coupled equations
	\begin{widetext}
		\begin{subequations}
			\label{eq:coupled_XV}
			\begin{align}
				\Omega X_{m,p'} &= m\overline{\omega}_s X_{m,p'} - m \Bigg[ i\kappa \sum_{p=-\infty}^{\infty}\frac{Z_{p,\ell}(m\overline{\omega}_s)}{\omega_{p,\ell}} F_{m,pp'} \sum_{m'\neq 0} X_{m',p} + \kappa_{\text{FB}}(m\overline{\omega}_s) N_{m,p'} \sum_{m'\neq 0} V_{m'} \Bigg], \label{eq:eigY} \\
				\Omega V_m &= m\overline{\omega}_s V_m - m \Bigg[ i\kappa \sum_{p=-\infty}^{\infty} \frac{Z_{p,\ell}(m\overline{\omega}_s)}{\omega_{p,\ell}} P_{m,p} \sum_{m'\neq 0} X_{m',p} + \kappa_{\text{FB}}(m\overline{\omega}_s) Q_m \sum_{m'\neq 0} V_{m'} \Bigg]. \label{eq:eigU}
			\end{align}
		\end{subequations}
	\end{widetext}
	
	With the definition of another bilinear product, applied to individual azimuthal functions $A_m$ and $B_m$, as
	\begin{equation}
		\mathcal{M}[A_m, B_m] := \int_0^\infty dJ \frac{\partial{\Psi_0}}{\partial J} A_m(J) B_m(J),
		\label{eq:M_functional}
	\end{equation}
	with the weighting function given by ${\partial{\Psi_0}}/{\partial J}$, the quantities introduced in Eqs.~\eqref{eq:coupled_XV} are:
	\begin{align}
		F_{m,pp'} &= \mathcal{M}[H_{m,p'}, H_{m,p}^*], \\
		N_{m,p'}  &= \mathcal{M}[\hat{f}_m, H_{m,p'}], \\
		P_{m,p}   &= \mathcal{M}[\hat{f}_{m}^*, H_{m,p}^*], \\
		Q_m       &= \mathcal{M}[\hat{f}_m, \hat{f}_{m}^*].
	\end{align}
	
	Equations~\eqref{eq:coupled_XV} can be cast as a block eigenvalue problem:
	\begin{equation}
		\Omega
		\begin{bmatrix}
			\mathbf{X}_{m} \\
			V_m
		\end{bmatrix}
		= \sum_{m'\neq 0}
		\begin{bmatrix}
			\mathbf{A}_{m,m'} & \mathbf{B}_{m,m'} \\
			\mathbf{C}_{m,m'} & D_{m,m'}
		\end{bmatrix}
		\begin{bmatrix}
			\mathbf{X}_{m'} \\
			V_{m'}
		\end{bmatrix},
		\label{eq:block_eigenvalue}
	\end{equation}
	where the $p$-th element of $\mathbf{X}_{m}$ is $X_{m,p}$ and the sub-blocks are given by
	\begin{subequations}
		\begin{align}
			A_{mm',pp'} &= m\overline{\omega}_s \delta_{mm'} \delta_{pp'} - i m \kappa \frac{Z_{p,\ell}(m\overline{\omega}_s)}{\omega_{p,\ell}} F_{m,pp'}, \label{eq:sub_A} \\
			B_{mm',p'}   &= -m \kappa_{\text{FB}}(m\overline{\omega}_s) N_{m,p'}, \label{eq:sub_B} \\
			C_{mm',p}     &= -i m \kappa \frac{Z_{p,\ell}(m\overline{\omega}_s)}{\omega_{p,\ell}} P_{m,p}, \label{eq:sub_C} \\
			D_{mm'}     &= m\overline{\omega}_s \delta_{mm'} - m \kappa_{\text{FB}}(m\overline{\omega}_s) Q_{m}. \label{eq:sub_D}
		\end{align}
		\label{eq:submatrices}
	\end{subequations}
	
	By truncating~$|m| \le m_{\max}$ and selecting the harmonics~$p$, the complex coherent frequencies $\Omega$ are the eigenvalues of the finite matrix problem in Eq.~\eqref{eq:block_eigenvalue}.
	
	In the absence of feedback, $\kappa_{\text{FB}} = 0$, the block $\mathbf{B}_{m,m'}$ vanishes, leaving $\Omega \mathbf{X}_m = \sum_{m'} \mathbf{A}_{m,m'} \mathbf{X}_{m'}$, which is an eigenvalue equation independent of the centroid $V_m$. This reproduces the effective synchrotron frequency model without feedback presented in Ref.~\cite{Alves-6384}.
	
	\subsection{Inclusion of Realistic Hardware Effects via Transfer Functions}\label{sec:tf_hardware}
	The formalism developed above is independent of the implementation of the feedback system. Any linear time-invariant controller can be incorporated by evaluating its transfer function at the coherent frequency $\Omega$. Different feedback types therefore modify the Lebedev equation only through $\kappa_{\text{FB}}(\omega)=\frac{2\pi ec}{E_{0}C_{0}}T_{\text{FB}}(\omega)$.
	
	Several practical feedback systems can be represented within this framework. The simplest case is a constant-gain controller,
	\begin{equation}
		T_{\text{FB}}(\omega)=G \label{eq:40}
	\end{equation}
	This model neglects realistic features such as bandwidth limitations and delays and corresponds to ideal proportional feedback. The gain $G$ may, in general, be complex, $G=G_{R}+iG_{I}$ where $G_{R}$ and $G_{I}$ are the real and imaginary parts of the feedback gain, respectively.
	
	These two components have different physical effects on the beam dynamics. The real part introduces a voltage in phase with the bunch centroid oscillation; it acts as reactive feedback, modifying the restoring force acting on the beam and shifting the coherent oscillation frequency.
	The imaginary part of the gain produces a voltage in quadrature with the centroid motion, acting as resistive feedback. Depending on its sign, it either removes energy from the coherent oscillation (damping), or feeds energy to the beam (anti-damping). Therefore, the imaginary part of the gain predominantly modifies the growth rate of the coherent motion.
	
	A more realistic model includes a finite delay inherent to the feedback system:
	\begin{equation}
		T_{\text{FB}}(\omega)=Ge^{i\omega\tau_{d}} \label{eq:41}
	\end{equation}
	where $\tau_{d}$ is the total latency between beam measurement and correction. The delay introduces a frequency-dependent phase shift that strongly influences the feedback performance.
	
	The finite bandwidth of the kicker cavity and power amplifier can be modeled as a single-pole low-pass response, in which the kicker field persists over consecutive bunches:
	
	\begin{equation}T_{\text{FB}}(\omega) = G \frac{1 - \alpha}{1 - \alpha e^{i 2\pi \ell / N_b}} \label{eq:tfb_crosstalk}\end{equation}
	\begin{equation}\alpha = e^{-2\pi \Delta f_{\mathrm{BW}} T_b}\label{eq:alpha_decay}\end{equation}
	where $\ell$ is the coupled-bunch mode index, $N_b$ is the total number of bunches ($N_b = M$ in the notation of Sec.~\ref{sec:theory}), $\Delta f_{\mathrm{BW}}$ is the 3-dB bandwidth of the feedback rf system, $T_b$ denotes the bunch spacing period, and $\alpha$ is the bunch-to-bunch attenuation factor.
	
	In general, any transfer function obtained from measurements or simulations can be directly incorporated into the present formalism through $T_{\text{FB}}(\omega)$ without modifying the derivation.
	
	The theoretical toolkit developed in this section---the Lebedev dispersion solver and the effective synchrotron frequency eigenvalue formulation, both extended to include BbB feedback---has been implemented in the open-source \texttt{pycolleff} package~\cite{pycolleff} and is used for all semi-analytical results in Sec.~\ref{sec:tracking}.
	
	\section{Comparative analysis with macroparticle tracking}\label{sec:tracking}
	
	\subsection{Simulation Setup and Parameters}
	To benchmark the analytical framework, macroparticle tracking simulations were carried out using the \texttt{STABLE} code~\cite{He-6289}. Originally developed for transient beam loading studies, \texttt{STABLE} was augmented with a dedicated bunch-by-bunch feedback module. On each turn, the module records the bunch centroid displacements, computes the corresponding reactive voltage kick, and applies it uniformly across the macroparticles within each bucket. The implementation details and the extended code package are publicly available in Ref.~\cite{fang_stable_feedback_2026}.
	
	The tracking model adopts the nominal parameters of the MAX~IV $3\text{ GeV}$ storage ring operating in a double-rf configuration. Passive third-harmonic cavities are tuned near the flat-potential condition to achieve bunch lengthening, while the main rf cavity is assumed to maintain a constant gap voltage, with its beam-loading effects neglected. The primary machine and rf parameters are summarized in Table~\ref{tab:maxiv_parameters}. In all tracking runs, a uniform fill over all $h = 176$ buckets is used, with \num{10000} macroparticles per bunch. Both radiation damping and quantum excitation are included, and short-range wakefields are neglected.
	
	\begin{table}[htbp]
		\centering
		\caption{Main storage ring and harmonic cavity parameters for MAX~IV (harmonic-cavity quantities are given per cavity)~\cite{CullinanAndersson-6290}.}
		\label{tab:maxiv_parameters}
		\begin{tabular}{lcc}
			\hline\hline
			Parameter & Symbol & Value \\
			\hline
			\multicolumn{3}{l}{\textit{Storage Ring Parameters}} \\
			Beam energy & $E_0$ & $3\text{ GeV}$ \\
			Circumference & $C_0$ & $528\text{ m}$ \\
			rf frequency & $f_{\text{rf}}$ & $99.931\text{ MHz}$ \\
			Harmonic number & $h$ & $176$ \\
			Momentum compaction & $\alpha_c$ & $3.06 \times 10^{-4}$ \\
			Energy loss per turn (bare lattice) & $U_0$ & $363.8\text{ keV}$ \\
			Natural energy spread & $\sigma_{\delta 0}$ & $7.69 \times 10^{-4}$ \\
			Radiation damping time & $\tau_{\mathrm{rad}}$ & $25.2\text{ ms}$ \\
			\hline
			\multicolumn{3}{l}{\textit{rf System Parameters}} \\
			Main cavity (MC) voltage & $V_{\text{rf}}$ & $1.0\text{ MV}$ \\
			HC harmonic & $n_{\text{hc}}$ & $3$ \\
			HC unloaded quality factor & $Q_0$ & $20800$ \\
			HC geometric factor & $(R/Q)$ & $132.21\ \Omega$ \\
			\hline\hline
		\end{tabular}
	\end{table}

	\subsection{Time-Domain Evolution of Bunch Centroids and Lengths}\label{sec:time_domain}
	
\begin{figure}[htbp]
		\centering
		\includegraphics[width=\columnwidth]{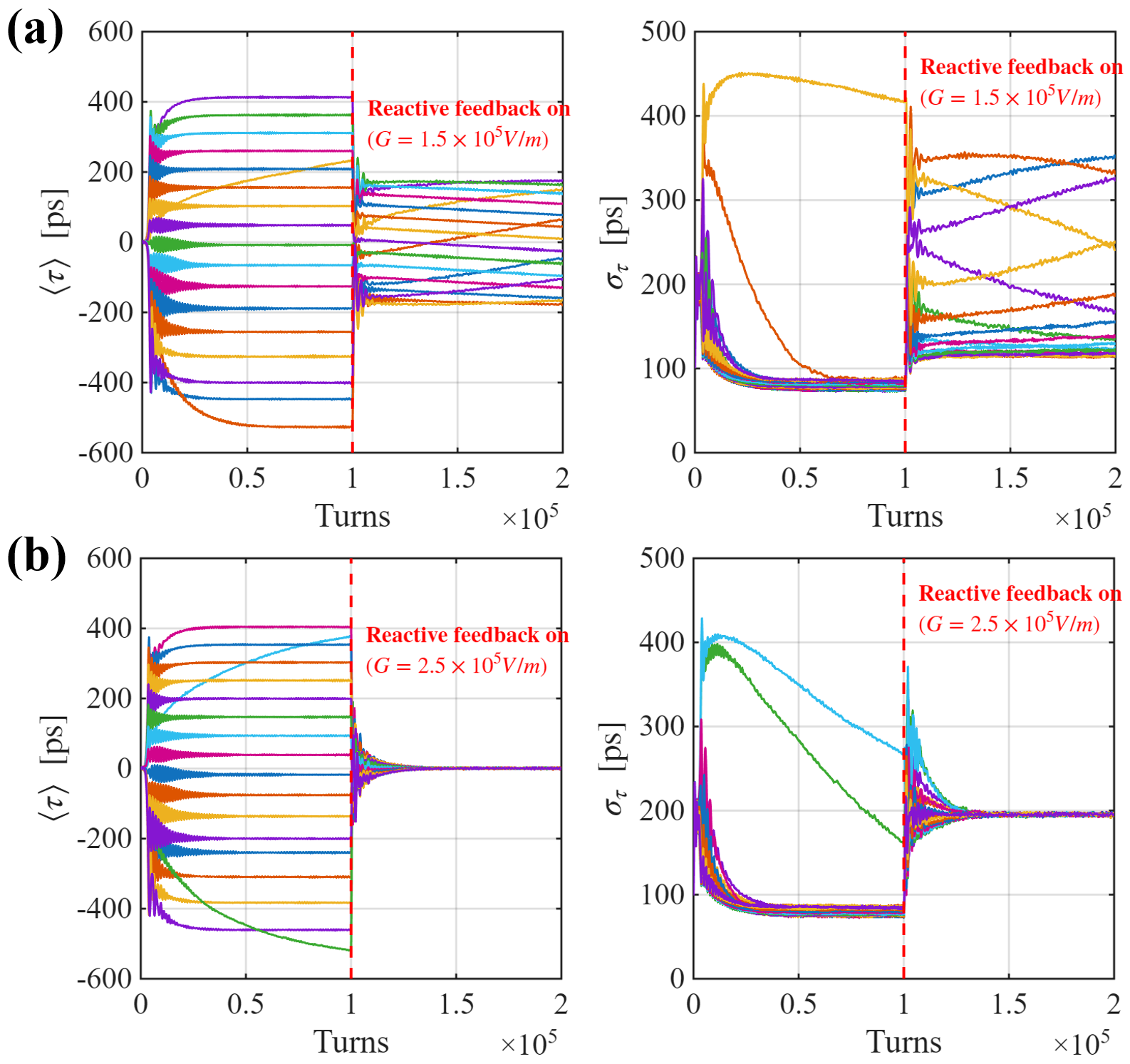}
		\caption{Temporal evolution of bunch centroid deviations $\langle \tau \rangle$ and rms bunch lengths $\sigma_\tau$ at $I_0 = 400\text{ mA}$, with reactive feedback engaged at turn $1\times 10^5$. (a) With an insufficient gain of $G = 1.5\times 10^5\ \mathrm{V/m}$, the PTBL effect is only partially mitigated, exhibiting residual coherent oscillations. (b) With a sufficient gain of $G = 2.5\times 10^5\ \mathrm{V/m}$, the PTBL oscillations are completely suppressed, and the bunches settle into a stable, stretched equilibrium state.}
		\label{fig:time_evolution}
	\end{figure}

	To intentionally induce the PTBL effect, the beam current of MAX~IV is set to $400\text{ mA}$ with three third-harmonic cavities operated for bunch lengthening, placing the machine in a regime beyond the PTBL stability boundary. After the instability fully develops, reactive feedback is turned on at turn $10^5$. To implement pure reactive feedback, the complex transfer function is set to a constant real gain, $T_{\text{FB}}(\Omega) \equiv G$, providing a voltage kick proportional to the instantaneous bunch centroid displacement.
	Figure~\ref{fig:time_evolution} compares the turn-by-turn evolution of the bunch centroid offsets $\langle \tau \rangle$ and rms bunch lengths $\sigma_\tau$ under different feedback gains $G$.
	
	During the initial $1\times 10^5$ turns, in which the feedback loop remains inactive, the periodic transient beam loading drives growing, large-amplitude, low-frequency oscillations in both the bunch centroid and the bunch length---the onset of the PTBL effect. At turn $1\times 10^5$, the reactive feedback is engaged. The subsequent beam dynamics are governed by the feedback gain $G$: when $G$ is below the critical gain threshold, the oscillation is partially mitigated and its frequency shifts upward, yet the residual PTBL modulation persists. Conversely, when the feedback gain $G$ exceeds the critical gain threshold, the coherent oscillations are completely damped out, and the bunches settle into a stable, stretched steady-state distribution with no residual coherent oscillation. This means that suppressing the PTBL effect requires the reactive feedback gain to exceed a well-defined threshold. The correction voltage demanded from the feedback system is small: with $G = 2.5\times 10^{5}\ \mathrm{V/m}$, the peak kick during the damping transient amounts to only about $300$~V.
	
	\subsection{Evaluation of Instability Threshold Product under Reactive Feedback}
	
\begin{figure}[htbp]
		\centering
		\includegraphics[width=1.0\columnwidth]{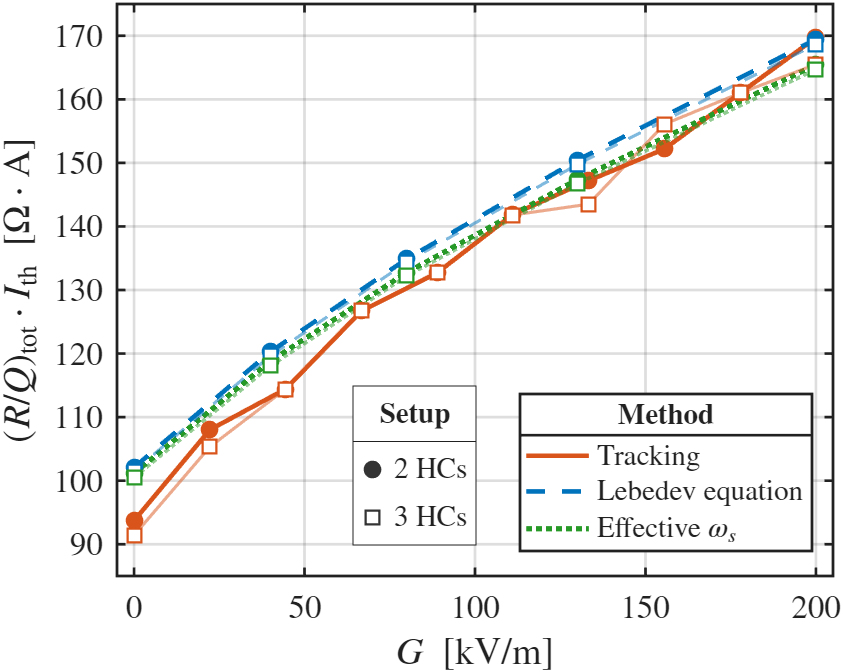}
		\caption{PTBL threshold, expressed as the impedance--current product $(R/Q)_{\mathrm{tot}} \cdot I_{\mathrm{th}}$, as a function of feedback gain $G$. Results from macroparticle tracking (solid lines) are compared with analytical evaluations using the Lebedev equation (dashed lines) and the effective $\omega_s$ approach (dotted lines) for systems with 2~HCs (solid circles) and 3~HCs (open squares).}
		\label{fig:gain_vs_current}
	\end{figure}

	As demonstrated in the preceding section, stabilizing the beam against the PTBL effect imposes a minimum requirement on the reactive feedback gain $G$. To systematically evaluate this stabilization boundary across the parameter space, we scan the feedback gain $G$ and determine the corresponding threshold of the impedance--current product, $(R/Q)_{\mathrm{tot}}\cdot I_{\mathrm{th}}$, where $(R/Q)_{\mathrm{tot}}$ is the total geometric factor of the harmonic cavities.
	Fig.~\ref{fig:gain_vs_current} compares the resulting stability boundaries predicted by the Lebedev equation, the effective synchrotron frequency model, and macroparticle tracking simulations using the \texttt{STABLE} code.
	
	For the semi-analytical models, the PTBL threshold is determined from the condition that the mode-1 growth rate remains below the radiation damping rate. In the \texttt{STABLE} tracking simulations, PTBL stability is evaluated using the centroid mode-1 amplitude, $a(t) = \big\vert{}h^{-1}\sum_{j=1}^{h}\bar{\tau}_{j}(t)\, e^{-2\pi i j/h}\big\vert{}$, where $\bar{\tau}_{j}(t)$ denotes the longitudinal centroid offset of bunch $j$ at turn $t$. Each configuration is tracked for $10^{6}$ turns and each configuration is classified as unstable if the median of $a(t)$ over the final $10000$ turns exceeds a threshold of $A_{\mathrm{dec}} = 1$~ps. This threshold is chosen an order of magnitude above the noise floor while remaining well below the nominal bunch length.
	
	As shown in Fig.~\ref{fig:gain_vs_current}, the thresholds obtained from the Lebedev equation, the effective synchrotron frequency method, and \texttt{STABLE} tracking agree well with one another, with a worst-case deviation of about $10\%$, confirming the consistency between the semi-analytical models and the tracking code.
	
	Across the examined gain range, the results for the 2-HC and 3-HC configurations overlap closely, confirming that the PTBL threshold scaling with the product $(R/Q) \cdot I$ that was identified in the HALF study \cite{HeLi-6145} remains valid in the presence of reactive feedback. Moreover, the threshold increases monotonically with $G$, indicating that a higher feedback gain provides stronger suppression against PTBL, consistent with the requirement for sufficient feedback gain $G$ established in the previous section.
	
	There are still differences among the three methods. The Lebedev equation yields a slightly higher threshold than the effective $\omega_s$ method because the latter neglects Landau damping induced by the synchrotron frequency spread. Accounting for this damping effect in the Lebedev equation moderately raises the threshold. On the other hand, the threshold from \texttt{STABLE} tracking is lower (by approximately 10\%) than the analytical predictions at low gains, but gradually approaches the theoretical curves as $G$ increases. Although a definitive explanation for this convergence requires further study, the discrepancy may reflect the sensitivity to the numerical instability criterion in tracking, as well as fine details of the bunch phase-space distribution.
	
	\subsection{Instability Threshold Current Versus Main Cavity Voltage}
	
	To investigate the parametric dependence and verify the agreement between theoretical models and tracking simulations, the instability threshold current $I_{\mathrm{th}}$ as a function of the main cavity rf voltage $V_{\mathrm{rf}}$ is examined across various reactive feedback gains $G$. The storage ring configuration is modeled with two passive harmonic cavities (rather than the three used in Sec.~\ref{sec:time_domain}), operating under the near-flat-potential condition.
	The main cavity voltage is scanned across the nominal operational range from $0.90$ to $1.10\ \mathrm{MV}$, while the reactive feedback gain $G$ is varied from $0$ to $3.0\times 10^5\ \mathrm{V/m}$.
	
\begin{figure}[htbp]
		\centering
		\includegraphics[width=\linewidth]{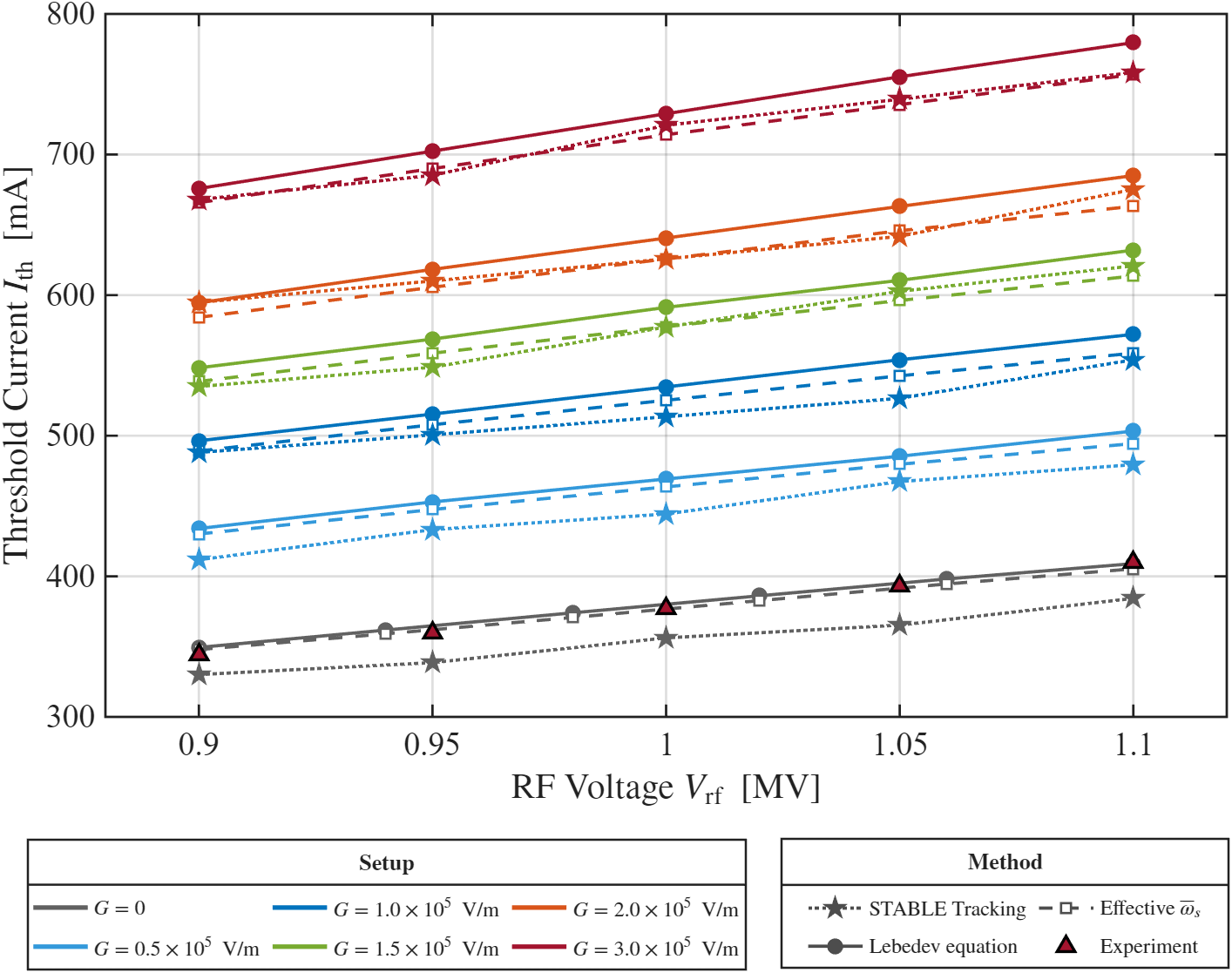}
		\caption{PTBL threshold beam current $I_{\mathrm{th}}$ as a function of the main rf cavity voltage $V_{\mathrm{rf}}$ across various reactive feedback gains ($G = 0$, $0.5\times 10^5$, $1.0\times 10^5$, $1.5\times 10^5$, $2.0\times 10^5$, and $3.0\times 10^5\ \mathrm{V/m}$). The semi-analytical predictions obtained from the Lebedev equation (solid lines with circles) and the effective synchrotron frequency approximation (dashed lines with squares) are benchmarked against macroparticle tracking simulations using the \texttt{STABLE} code (dotted lines with stars) and against experimental data from MAX IV studies without feedback~\cite{CullinanAndersson-6290} (black triangles).}
		\label{fig:vrf_vs_ith}
	\end{figure}

	As shown in Fig.~\ref{fig:vrf_vs_ith}, the threshold beam current $I_{\mathrm{th}}$ increases monotonically with the main cavity voltage $V_{\mathrm{rf}}$ across all configurations, in good agreement with the scaling trend reported in Refs.~\cite{HeLi-6145,Alves-6384}.
	Consistent with the findings in the previous subsection, the semi-analytical predictions from the Lebedev equation and the effective synchrotron frequency approximation match each other closely, while slightly exceeding the macroparticle tracking results from \texttt{STABLE}. Overall, the quantitative agreement among the theoretical models, tracking simulations, and the experimental baseline at $G = 0$ corroborates the calculation framework.
	
	More importantly, introducing reactive feedback substantially raises the PTBL threshold current across the entire voltage range. At $G = 3.0\times 10^5\ \mathrm{V/m}$, $I_{\mathrm{th}}$ is enhanced by nearly a factor of two compared to the feedback-free baseline ($G = 0\ \mathrm{V/m}$). While raising the main cavity voltage also raises the current threshold, it does so at the cost of a substantial increase in main rf power consumption. Reactive feedback avoids this cost: it suppresses the PTBL effect with only the correction voltage of an additional feedback cavity, allowing the main cavity to operate at a lower voltage.
	
	\subsection{Engineering Constraints and Operational Boundaries}
	
	The preceding analysis and simulations were carried out under the assumption of an idealized bunch-by-bunch feedback system. In realistic implementations, however, physical constraints and hardware non-idealities are inevitable. These non-idealities cause part of the reactive feedback signal to leak into other oscillation modes, potentially driving previously stable modes unstable.
	
	In this subsection, we systematically evaluate two critical hardware limitations. The first is the total feedback loop latency---originating from the physical signal path and digital processing time between the beam position pickup and the rf kicker. This delay introduces phase lag that leaks an anti-damping component into the dipole motion. The second is the finite bandwidth of the rf system, consisting of both the solid-state power amplifier and the rf kicker cavity. Because each individual bunch demands an independent kick, restricted system bandwidth prevents the cavity fields from responding instantaneously between adjacent buckets, thereby causing inter-bunch crosstalk and potentially exciting higher-order coupled-bunch instabilities.
	
	\subsubsection{Impact of feedback latency: Excitation of mode-0 instability}
	
	We consider the MAX IV storage ring operating at $400\ \mathrm{mA}$ with three passive harmonic cavities tuned to the near-flat-potential condition, and a reactive feedback gain of $G = 2.5 \times 10^5\ \mathrm{V/m}$.
	To evaluate the impact of latency, an overall feedback loop delay of 20 turns ($\tau_d = 20\,T_0 \approx 35\ \mu\mathrm{s}$) is introduced into the feedback path.
	
\begin{figure}[htbp]
		\centering
		\includegraphics[width=\columnwidth]{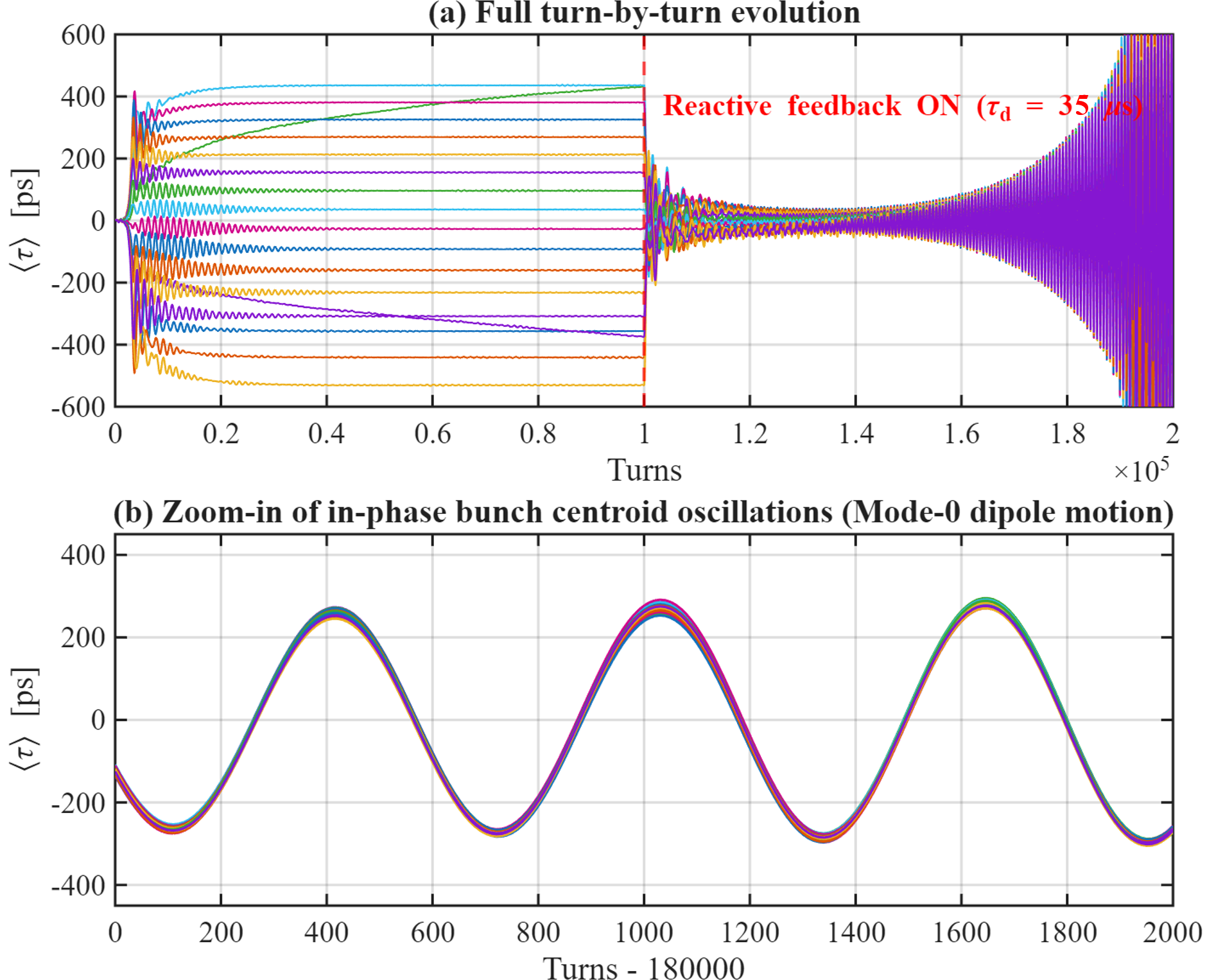}
		\caption{Turn-by-turn evolution of bunch centroid time offsets $\langle\tau\rangle$ simulated with the \texttt{STABLE} code at $400\ \mathrm{mA}$ with three harmonic cavities tuned to the near-flat-potential condition. (a) Long-term evolution over $2.0\times 10^5$ turns, where the reactive feedback ($G = 2.5\times 10^5\ \mathrm{V/m}$) with a 20-turn delay ($\tau_d \approx 35\ \mu\mathrm{s}$) is engaged at the $1.0\times 10^5$-th turn (red dashed line). The initial PTBL effect is promptly suppressed, but is followed by exponential growth of a dipole oscillation. (b) Zoomed-in view of the centroid motion between turns $1.80\times 10^5$ and $1.82\times 10^5$, demonstrating that all bunches oscillate in phase with uniform amplitude, characteristic of the mode-0 ($\ell = 0$) dipole instability.}
		\label{fig:latency_mode0}
	\end{figure}

	Macroparticle tracking simulations are performed using the \texttt{STABLE} code, where the reactive feedback is activated at the $10^5$-th turn after the PTBL effect has fully developed. Figure~\ref{fig:latency_mode0} illustrates the turn-by-turn evolution of the bunch centroid positions obtained from the tracking simulation.
	As observed in the figure, the PTBL effect is rapidly suppressed once the reactive feedback is engaged. However, over subsequent turns, a coherent dipole oscillation emerges and exhibits pronounced exponential growth. A detailed inspection of the centroid distributions across the bunch train reveals that all bunches oscillate in phase with identical amplitudes, which is characteristic of the mode-0 ($\ell = 0$) instability.
	
	As formulated in Sec.~\ref{sec:tf_hardware}, loop latency can be incorporated into the theoretical framework by replacing the constant gain with a frequency-dependent transfer function, $T_{\mathrm{FB}}(\Omega) = G e^{i\Omega \tau_d}$. Substituting the feedback parameters into the Lebedev equation and the effective synchrotron frequency model confirms that the growth rate of mode-0 increases significantly, surpassing the radiation damping rate. Furthermore, as summarized in Table~\ref{tab:mode0_comparison}, both the mode-0 coherent oscillation frequencies and the growth rates predicted by the theoretical models agree well with the tracking simulations, validating both the analytical formulations in Sec.~\ref{sec:tf_hardware} and the tracking model.
	
	Physically, this mode-0 instability arises because the time delay $\tau_d$ shifts the phase of the kick relative to the bunch displacement by $\Delta \phi \approx \Omega \tau_d$. Consequently, the feedback signal no longer provides purely reactive focusing; instead, it develops a resistive component proportional to $G\sin(\Omega \tau_d)$. This parasitic resistive term acts as an anti-damping force that continuously pumps energy into the coherent dipole motion. Because all bunches experience identical phase lags under a constant global latency, mode-0---in which all bunches oscillate in phase---receives the strongest coherent excitation and becomes unstable once the anti-damping rate exceeds radiation damping.
	
\begin{figure}[tb]
		\centering
		\includegraphics[width=\linewidth]{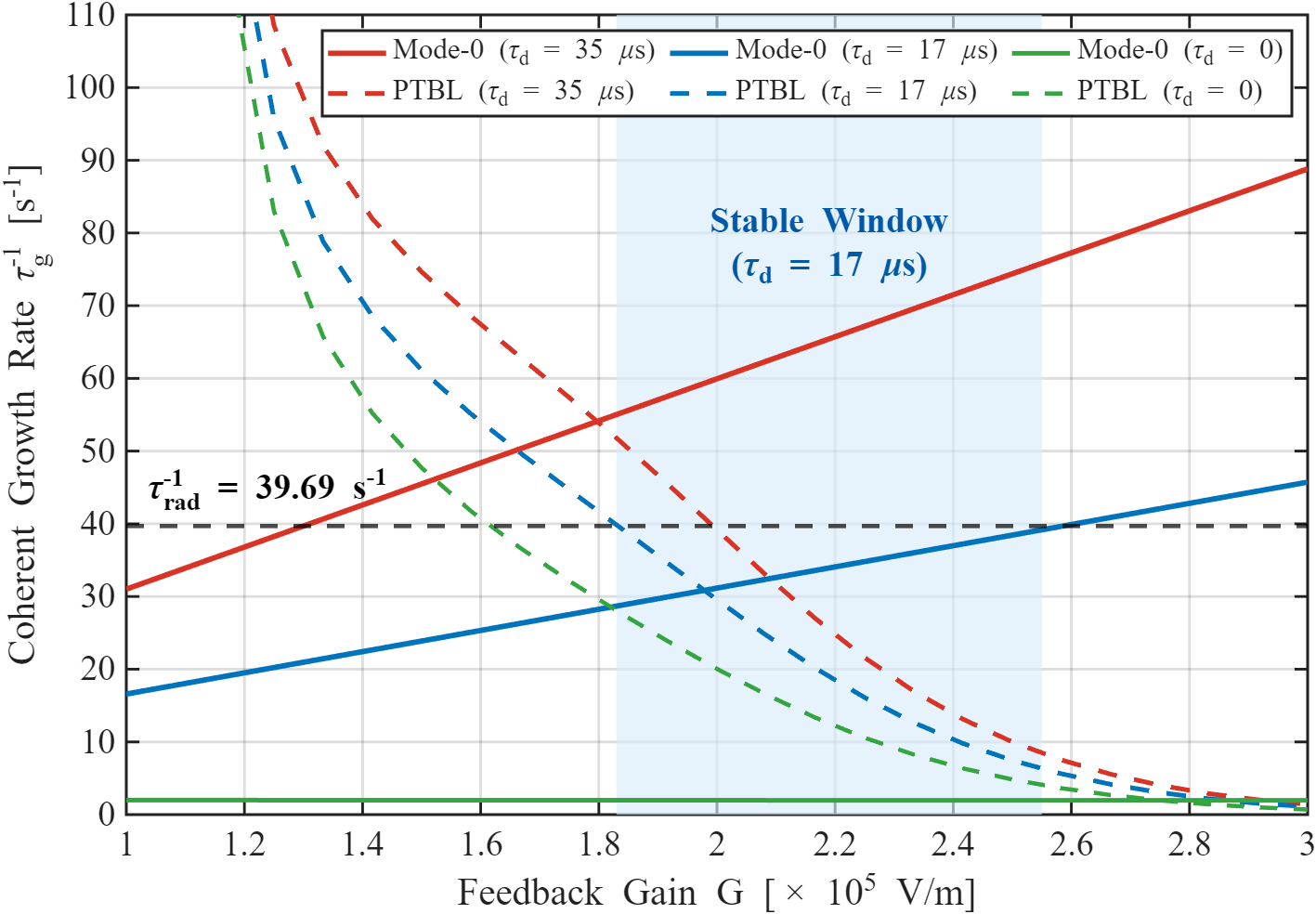}
		\caption{Calculated growth rates of the PTBL effect (dashed lines) and the mode-0 dipole instability (solid lines) as a function of the reactive feedback gain $G$ for various loop latencies ($\tau_d = 0$, $17\ \mu\text{s}$, and $35\ \mu\text{s}$). The horizontal black dashed line indicates the natural radiation damping rate ($\tau_{\text{rad}}^{-1} = 39.69\ \text{s}^{-1}$). For $\tau_d = 17\ \mu\text{s}$, a distinct stable operational window (shaded region, $1.83\times 10^5 \le G \le 2.55\times 10^5\ \text{V/m}$) exists where both growth rates fall below the radiation damping threshold. In contrast, at $\tau_d = 35\ \mu\text{s}$, the stable window completely closes.}
		\label{fig:latency_growth_rate_scan}
	\end{figure}

	\begin{table}[htbp]
		\centering
		\caption{Comparison of the mode-0 coherent oscillation frequency $f_{\text{coh}}$ and coherent growth rate $\tau_g^{-1}$ obtained from semi-analytical models and \texttt{STABLE} tracking. Calculations are performed for the MAX IV storage ring at $I_0 = 400\ \text{mA}$ with three harmonic cavities operating in the near-flat-potential condition, subjected to a reactive feedback gain of $G = 2.5\times 10^5\ \text{V/m}$ and a loop latency of $\tau_d = 35\ \mu\text{s}$ (20 turns).}
		\label{tab:mode0_comparison}
		\begin{tabular}{lcc}
			\hline\hline
			Method & $f_{\text{coh}}$ [$\text{Hz}$] & $\tau_g^{-1}$ [$\text{s}^{-1}$] \\
			\hline
			Full Lebedev                    & $979.41$ & $74.39$ \\
			Effective $\overline{\omega}_s$ & $976.21$ & $73.67$ \\
			STABLE Tracking                 & $925.29$ & $76.13$ \\
			\hline\hline
		\end{tabular}
	\end{table}
	To investigate the dependence of instability growth rates on loop latency, a parameter scan is performed over various feedback gains $G$ and three values of the latency $\tau_d$, as shown in Fig.~\ref{fig:latency_growth_rate_scan}. As $G$ increases, the PTBL growth rate is progressively suppressed, whereas that of the mode-0 instability rises monotonically---an escalation that becomes significantly steeper at larger latencies $\tau_d$.
	
	Consequently, if the loop latency is excessively long, the mode-0 growth rate exceeds the radiation damping limit before the PTBL effect can be effectively extinguished. This imposes an upper bound on the feedback gain $G$, as exceeding it excites the mode-0 instability. To preserve an operable gain window, a strict constraint on the loop latency is thus required. For the MAX IV parameters considered here, maintaining the overall loop latency within $17\ \mu\mathrm{s}$ (approximately 10 turns) is sufficient to ensure stable operation, as it leaves an operational gain window. Within this range, choosing an intermediate gain such as $G = 2.0\times 10^5\ \mathrm{V/m}$ suppresses both the PTBL effect and the mode-0 instability.
	
	\subsubsection{Impact of feedback bandwidth: Excitation of higher-order mode instability}
	
	Following the baseline configuration established in the preceding subsection (MAX IV at $400\ \mathrm{mA}$ with three harmonic cavities and $G = 2.5 \times 10^5\ \mathrm{V/m}$), the same ring parameters and optimal harmonic cavity tuning are maintained.
	The reactive feedback is similarly engaged at the $10^5$-th turn, while the loop latency is set to zero and the 3-dB bandwidth of the rf feedback system is chosen as $\Delta f_{\mathrm{BW}} = 20\ \mathrm{MHz}$.
	
\begin{figure}[htbp]
		\centering
		\includegraphics[width=\columnwidth]{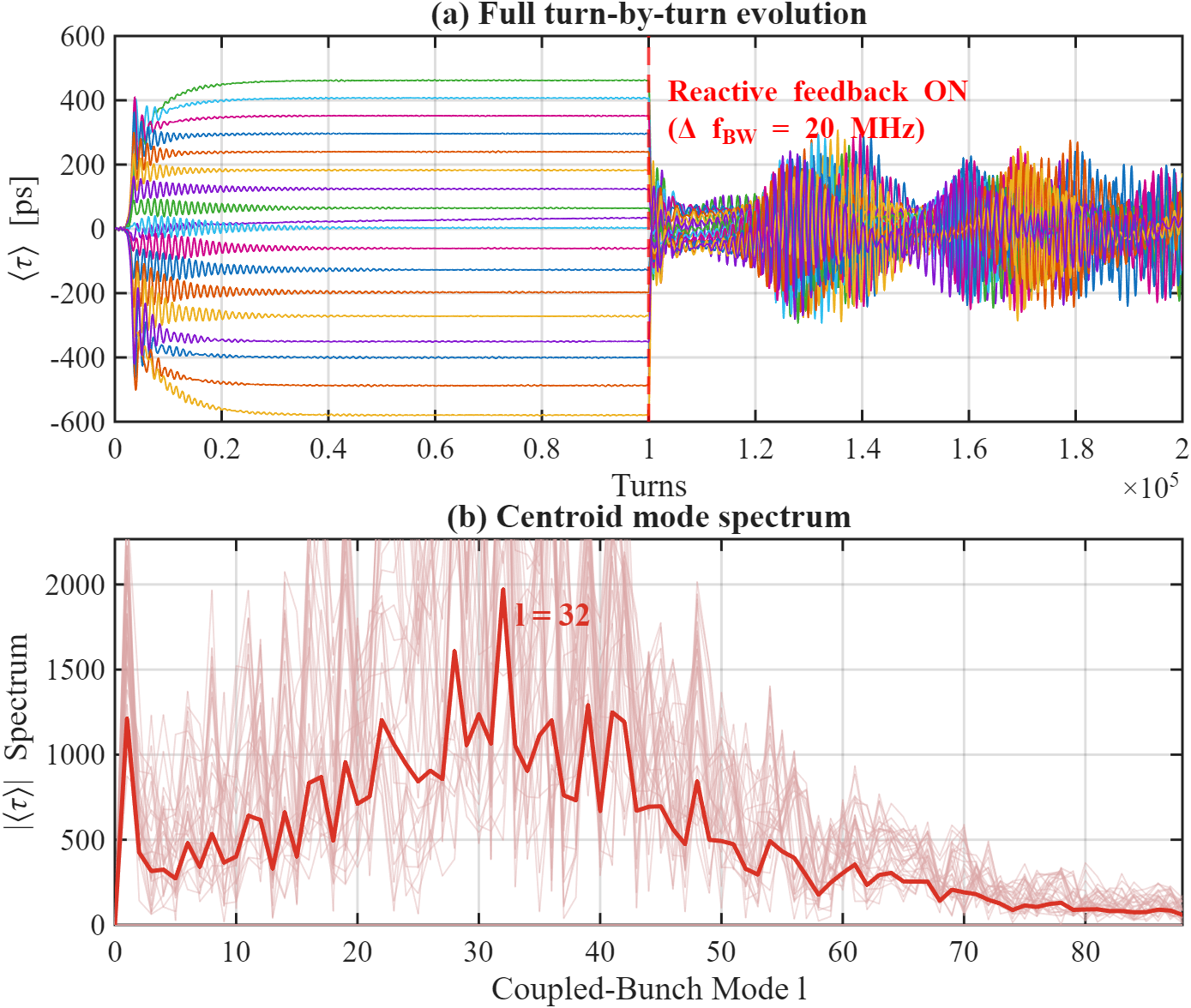}
		\caption{Excitation of higher-order coupled-bunch instability induced by finite rf system bandwidth ($\Delta f_{\mathrm{BW}} = 20\ \text{MHz}$, $\tau_d = 0$, $G = 2.5\times 10^5\ \mathrm{V/m}$). (a) Turn-by-turn bunch centroid evolution over $2.0\times 10^5$ turns. The reactive feedback is activated at turn $1.0\times 10^5$ (red dashed line), mitigating the PTBL pattern but triggering severe multi-bunch beating oscillations. (b) Spatial Fourier spectrum of the bunch centroids evaluated over turns $1.70\times 10^5$ to $1.71\times 10^5$, where light red traces denote instantaneous spectra and the bold red curve shows the ensemble average. The pronounced peak at $\ell \approx 32$ confirms the selective destabilization of higher-order coupled-bunch modes.}
		\label{fig:bandwidth_mode_spectrum}
	\end{figure}

	The tracking results from the \texttt{STABLE} code are shown in Fig.~\ref{fig:bandwidth_mode_spectrum}, which illustrates the turn-by-turn bunch centroid evolution and the corresponding spatial mode spectrum. As shown in Fig.~\ref{fig:bandwidth_mode_spectrum}(a), once the reactive feedback is turned on at the $10^5$-th turn, the original PTBL effect is quickly suppressed. However, it is replaced by strong centroid oscillations that vary from bunch to bunch. The spatial Fourier transform across the bunch train shows a clear peak near mode $\ell \approx 32$ [Fig.~\ref{fig:bandwidth_mode_spectrum}(b)], indicating that a higher-order coupled-bunch instability is excited.
	
\begin{figure}[tb]
		\centering
		\includegraphics[width=\columnwidth]{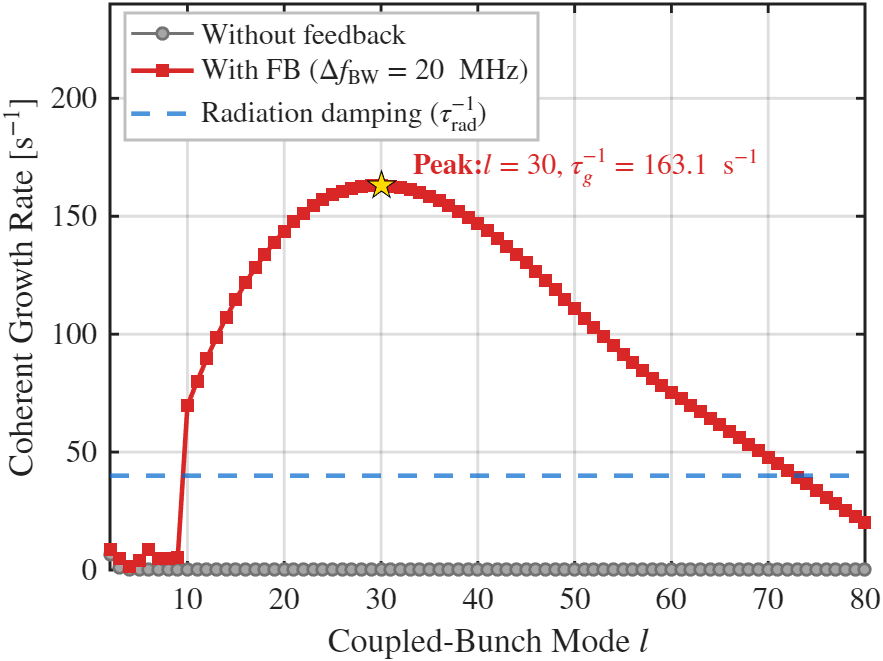}
		\caption{Calculated coherent growth rates for higher-order coupled-bunch modes ($\ell \ge 2$) with and without reactive feedback, assuming a feedback bandwidth of $\Delta f_{\mathrm{BW}} = 20\ \text{MHz}$ and a gain of $G = 2.5 \times 10^5\ \text{V/m}$. The dashed horizontal line indicates the natural radiation damping rate ($\tau_{\mathrm{rad}}^{-1}$). While all higher-order modes remain quiescent in the open-loop state, the bunch-to-bunch memory effect under limited cavity bandwidth drives an unstable band across $\ell \approx 10$--$72$ that exceeds radiation damping, peaking at $\ell = 30$ with a growth rate of $163.1\ \text{s}^{-1}$.}
		\label{fig:higher_order_growth}
	\end{figure}

	To compare these tracking results with theoretical predictions, the finite cavity bandwidth is incorporated using the transfer function derived in Sec.~\ref{sec:tf_hardware} [Eq.~\eqref{eq:tfb_crosstalk}]. The coherent growth rates across all coupled-bunch modes are calculated using the modified Lebedev equation, as shown in Fig.~\ref{fig:higher_order_growth}. Finite cavity bandwidth excites an unstable band ($\ell = 10$--$72$) exceeding the radiation damping threshold, whereas these modes remain completely damped without feedback. The growth rate envelope peaks at mode $\ell = 30$, which agrees well with the dominant mode observed in the \texttt{STABLE} tracking simulations and confirms the consistency between the semi-analytical framework and numerical tracking.
	
	Physically, the instability arises from the inter-bunch memory effect caused by the finite response time of the rf system. When the bandwidth $\Delta f_{\mathrm{BW}}$ is insufficient, the field induced to correct a given bunch does not decay before subsequent bunches arrive ($\tau_{\mathrm{cav}} \gtrsim T_b$, with $\tau_{\mathrm{cav}} = (2\pi\Delta f_{\mathrm{BW}})^{-1}$ the field decay time). Because coupled-bunch mode $\ell$ imposes a spatial phase advance $\Delta\theta_{\ell} = 2\pi \ell / N_b$ between adjacent bunches ($N_b = h$ for the uniform fill considered here), this lingering wakefield acts as an unintended crosstalk kick on trailing bunches.
	For low-order modes ($\ell \approx 0, 1$), adjacent bunches oscillate nearly in phase ($\Delta\theta_{\ell} \approx 0$), preserving the in-phase restoring nature of the reactive kick. However, for higher-order modes, the large inter-bunch phase difference causes the leaked field to project onto the bunch velocity rather than its displacement. This effectively distorts the reactive feedback into a resistive, anti-damping force, selectively driving modes around $\ell \approx 30$ where this phase misalignment peaks.
	
\begin{figure}[tb]
		\centering
		\includegraphics[width=\columnwidth]{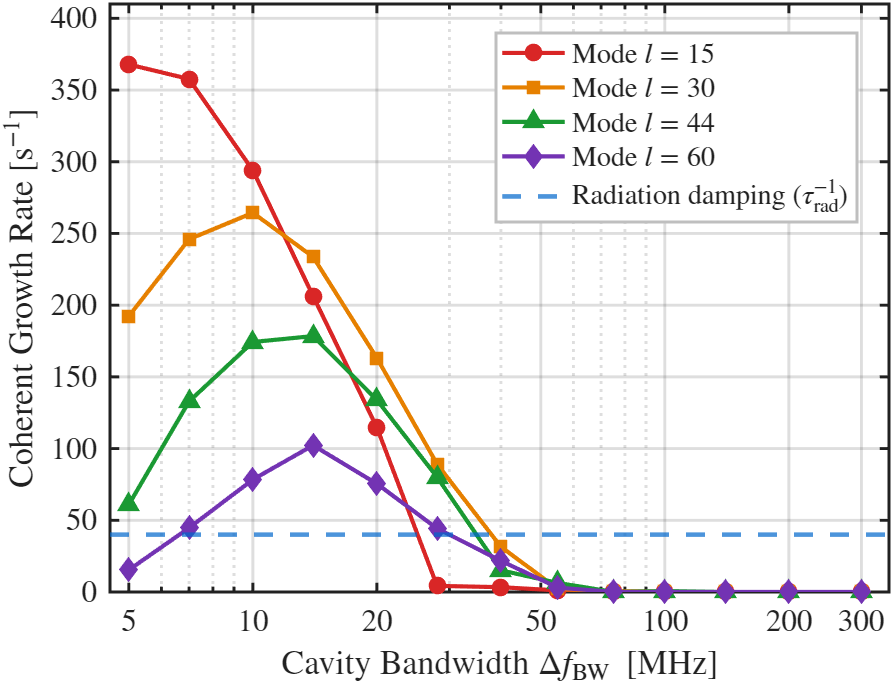}
		\caption{Coherent growth rates of representative higher-order coupled-bunch modes ($\ell = 15, 30, 44, 60$) as a function of the rf cavity bandwidth $\Delta f_{\mathrm{BW}}$, calculated at a feedback gain of $G = 2.5 \times 10^5\ \text{V/m}$. The horizontal dashed line denotes the natural radiation damping rate ($\tau_{\mathrm{rad}}^{-1} = 39.7\ \text{s}^{-1}$). As the bandwidth expands, inter-bunch wakefield memory rapidly decays, suppressing all higher-order modes below the radiation damping threshold once $\Delta f_{\mathrm{BW}} \gtrsim 50$~MHz.}
		\label{fig:growth_vs_bandwidth}
	\end{figure}

	To quantify the required feedback bandwidth, we perform a theoretical bandwidth scan and examine the coherent growth rates of representative higher-order modes as a function of $\Delta f_{\mathrm{BW}}$, as illustrated in Fig.~\ref{fig:growth_vs_bandwidth}. Overall, widening the bandwidth suppresses the growth rates by accelerating the decay of the residual kicker field between consecutive bunches.
	In the narrow-bandwidth regime ($5$--$15$~MHz), the growth rates of the larger-$\ell$ modes first increase with bandwidth before decreasing. This occurs because moderate bandwidths tailor the field decay time to match the inter-bunch phase advance, producing the strongest anti-damping projection.
	Beyond this peak, further expanding the bandwidth ensures that the residual field largely clears within the bunch spacing, leading to a steep reduction in growth rates. To prevent the excitation of higher-order instabilities, an explicit lower bound on the feedback bandwidth must be imposed. For our parameters, maintaining $\Delta f_{\mathrm{BW}} \ge 50$~MHz ensures that all higher-order modes are damped below the natural radiation damping baseline. This specification is readily satisfied by existing bunch-by-bunch longitudinal feedback hardware: the kicker cavity at MAX~IV, for example, provides a $3\text{-dB}$ bandwidth of approximately $69$~MHz~\cite{Ollson:2017dpu}, demonstrating that the proposed scheme can be accommodated with current infrastructure.
	
	\subsection{Extension to Triple-rf Systems}
	\label{sec:multi_rf_extension}
	
	In the previous subsections, we evaluated the double-rf system of MAX~IV with the Lebedev equation method, the effective synchrotron frequency method, and \texttt{STABLE} tracking simulations, and showed that reactive feedback effectively suppresses the PTBL effect.
	In fact, the reactive feedback scheme is also applicable to triple-rf and higher multi-rf systems. This can be understood from the time-domain mechanism: the PTBL effect is driven by a positive feedback loop, where the lengthened bunch has a very weak restoring force and is easily perturbed by voltage fluctuations, which in turn modulate the beam-induced voltage. Reactive feedback applies a voltage kick in phase with the centroid displacement, directly counteracting this perturbation and breaking the positive feedback loop. This suppression mechanism remains valid in multi-rf systems, regardless of the specific harmonic numbers of the cavities.
	
\begin{figure}[tb]
		\centering
		\includegraphics[width=\columnwidth]{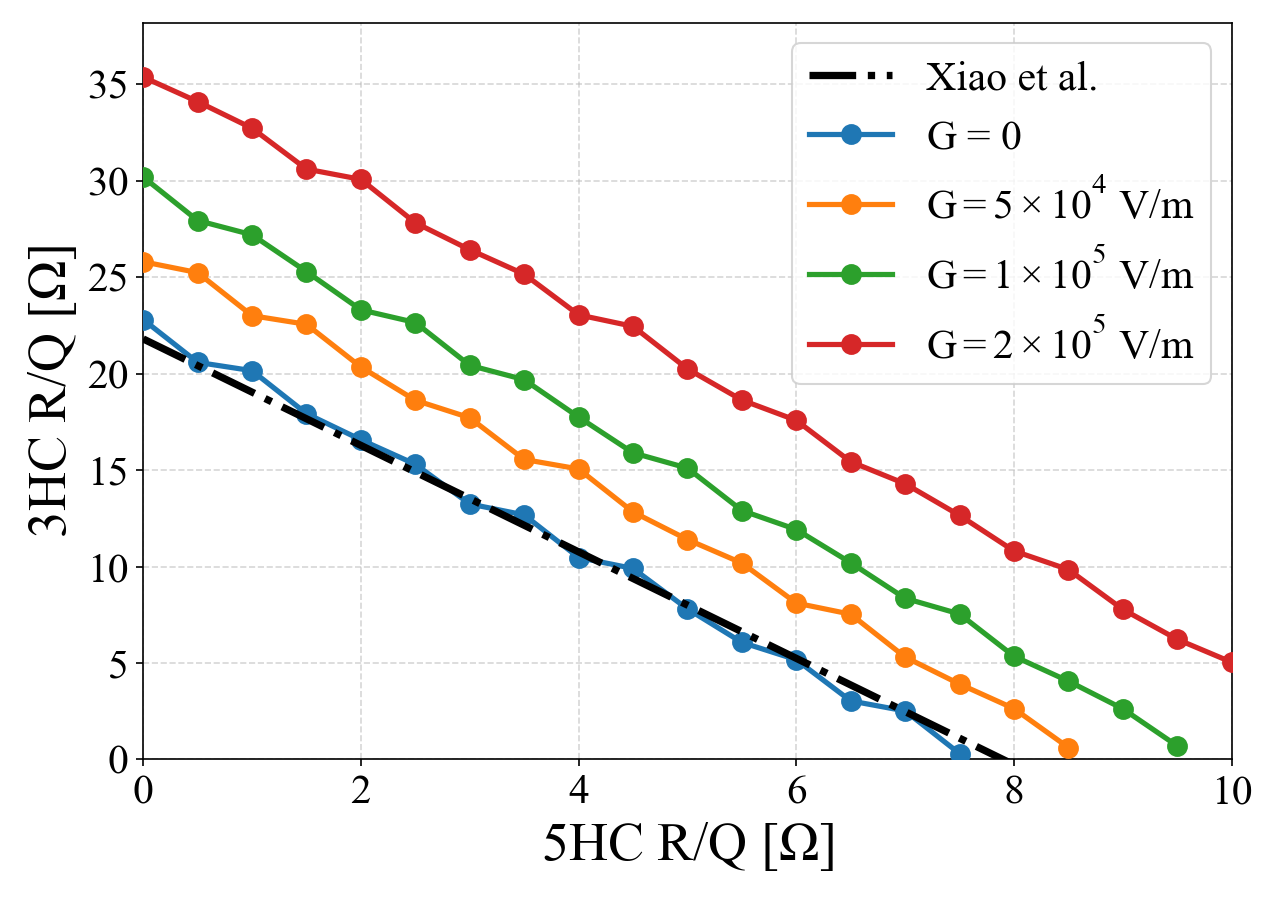}
		\caption{PTBL threshold boundaries in the $(R/Q)_{\mathrm{5HC}}$ versus $(R/Q)_{\mathrm{3HC}}$ geometric impedance plane for the HALF triple-rf system under the optimum bunch-lengthening condition. The dashed-dotted black line denotes the macroparticle tracking benchmark from Xiao et al.~\cite{xiao2025periodic}, while the solid colored lines with markers represent the semi-analytical boundaries calculated via the effective synchrotron frequency method across different reactive feedback gains ($G = 0$, $5\times 10^4$, $1\times 10^5$, and $2\times 10^5\text{ V/m}$). The region below each curve denotes the stable operating parameter domain.}
		\label{fig:triple_rf_thresholds}
	\end{figure}

	To verify this mitigation mechanism in multi-rf systems, we analyzed the triple-rf system of the HALF storage ring using the parameters from Xiao et al.~\cite{xiao2025periodic} as an example. The system consists of a main cavity, a third-harmonic cavity, and a fifth-harmonic cavity, operated at the optimum bunch-lengthening condition with a bunch length of 64.5~ps.
	Using the effective synchrotron frequency method described in Sec.~\ref{sec:theory}, we calculated the threshold curves in the $(R/Q)_{\text{5HC}}$ versus $(R/Q)_{\text{3HC}}$ plane for different reactive feedback gains ($G = 0$, $5\times 10^4$, $1\times 10^5$, and $2\times 10^5\text{ V/m}$).
	The resulting threshold curves are shown in Fig.~\ref{fig:triple_rf_thresholds}.
	
	When $G = 0$ (no reactive feedback), the calculated threshold boundary agrees with the macroparticle tracking results from Xiao et al.~\cite{xiao2025periodic}, confirming the accuracy of the effective synchrotron frequency method in triple-rf systems.
	As the gain $G$ increases, the threshold boundary shifts toward higher $R/Q$ values while maintaining the same slope. This confirms that reactive feedback suppresses the PTBL effect in the HALF triple-rf system, with larger gains providing wider stability margins.
	
	However, practical implementation imposes stringent demands on the feedback system. For the nominal design parameters of the HALF storage ring, where $(R/Q)_{\mathrm{3HC}} = 45\ \Omega$ and $(R/Q)_{\mathrm{5HC}} = 40\ \Omega$~\cite{xiao2025periodic}, extrapolating the threshold curves beyond the gain range shown in Fig.~\ref{fig:triple_rf_thresholds} indicates that suppressing the PTBL effect in the triple-rf system requires a reactive feedback gain of $G \approx 1.4\times 10^7\ \mathrm{V/m}$. This gain is nearly two orders of magnitude higher than that required for the double-rf configuration. Such an elevated gain implies substantially higher kick voltage demands and rf power requirements for the feedback kickers and solid-state power amplifiers, which may exceed typical operational limits in practical installations.
	
	These results indicate that reactive feedback remains effective in principle for triple-rf systems. Within the gain range studied, feedback raises the tolerable higher-harmonic $R/Q$ (equivalently, the PTBL threshold current at fixed $R/Q$), which enlarges the stability margin for extreme bunch lengthening.
	
	\section{Summary}
	\label{sec:summary}
	
	In this work, the suppression of the PTBL effect by reactive feedback has been investigated using theoretical analysis and macroparticle tracking simulations. The main stabilization mechanism of the feedback consists in providing additional coherent focusing for the coupled-bunch mode-1, shifting the coherent oscillation frequency away from zero and preventing the PTBL effect from being triggered.
	
	
	On the theoretical side, a feedback term was incorporated into the linearized Vlasov equation, extending the Lebedev equation and the effective synchrotron frequency model to include arbitrary feedback transfer functions. The resulting formulation enables the calculation of coherent frequencies and growth rates for any coupled-bunch modes in the presence of a feedback system. The formulation has been implemented in the open-source \texttt{pycolleff} package, expanding its capabilities for beam-stability studies with feedback. On the simulation side, a reactive-feedback module was implemented in the \texttt{STABLE} macroparticle tracking code and benchmarked using the parameters of the MAX IV storage ring. The tracking results reproduce the theoretical predictions: reactive feedback suppresses the PTBL oscillations above a threshold gain, significantly increasing the PTBL threshold, with good agreement between the semi-analytical calculations and tracking simulations over the investigated beam-current and $R/Q$ ranges.
	
	The analysis also establishes practical constraints on the parameters of the feedback system. Loop latency can destabilize the coupled-bunch mode-0, and low feedback bandwidth can excite higher-order coupled-bunch modes. Together, these constraints define specifications of maximum loop latency, minimum bandwidth, and an operational gain window for the feedback system. For the MAX IV parameters at 400 mA with three harmonic cavities, the reactive feedback system requires a loop latency below approximately \SI{17}{\micro\second} and a bandwidth above $\SI{50}{\mega\hertz}$ for stable operation.
	
	The applicability of the scheme to triple-rf systems was briefly investigated using the proposed configuration of the HALF storage ring. The semi-analytical results indicate that reactive feedback can also alleviate the PTBL effect in triple-rf systems, although a higher threshold gain is required than for the double-rf case. The feedback helps to extend the stable operating region in the $(R/Q)_{\mathrm{3HC}}$--$(R/Q)_{\mathrm{5HC}}$ parameter space while preserving the characteristic linear dependence between the two harmonic cavity $R/Q$ values.
	
	The combination of the frequency-domain semi-analytical model and time-domain tracking simulations developed in this work provides a framework for analyzing reactive feedback systems in electron storage rings with harmonic cavities. Experimental implementation and testing remain necessary to validate the predicted performance and establish the practical effectiveness of these systems in suppressing the PTBL effect.
	
	\bibliography{bi}
	
\end{document}